\PassOptionsToPackage{hyphens}{url}
\documentclass[sigconf,screen,nonacm]{acmart}

\usepackage{booktabs,array,calc,graphicx,xcolor}
\usepackage{xurl}
\newcommand{\bb}[1]{\boldsymbol{#1}}

\theoremstyle{remark}
\newtheorem*{remark}{Remark}

\graphicspath{{figures/}}
\title[Locational Pricing for Generative-AI Services via Token-Flow Market Clearing]{Locational Pricing for Generative-AI Services \texorpdfstring{\\}{ } via Token-Flow Market Clearing}
\author{Shaohui Liu}
\affiliation{%
  \institution{Massachusetts Institute of Technology}
  \city{Cambridge}
  \state{Massachusetts}
  \country{USA}
}
\email{shaohuil@mit.edu}
\renewcommand{\shortauthors}{Shaohui Liu}

\begin{abstract}
GenAI services are in an early yet fast expanding phase. Providers compete on model capability and service quality, while the underlying infrastructure remains expensive and heterogeneous across regions, workloads, and compute assets. If these services diffuse into routine daily use, the relevant engineering problem becomes not only better models but also efficient dispatch on a geographically distributed AI service infrastructure. To address this, we formulate a network-constrained token-flow market that clears AI workloads across compute nodes and communication links. The baseline model is a linear program that co-optimizes routing and processing subject to compute-capacity and bandwidth constraints; its dual variables define location- and workload-specific marginal service prices. We further introduce a transfer-aware extension that prices data movement in physical units and isolates bandwidth congestion rents. In a 5-node U.S. case study, the transfer-aware model uncovers four saturated backbone links and raises total operating cost by 2.7\% relative to the token-equivalent baseline, while tightening the chatbot latency limit from 100~ms to 15~ms increases one locational price by 117\%. A 20-node scale-up exhibits the same merit-order dispatch logic and becomes infeasible once demand exceeds aggregate capacity. These results suggest that locational pricing is a useful organizing principle for operating an emerging AI service infrastructure and, over time, for designing competitive markets around it.
\end{abstract}


\copyrightyear{2026}
\acmYear{2026}
\setcopyright{cc}
\setcctype{by-nc-nd}
\acmConference[ACM Sustainability Week Companion '26]{ACM Sustainability Week 2026}{June 22--25, 2026}{Banff, AB, Canada}
\acmBooktitle{ACM Sustainability Week 2026 (ACM Sustainability Week Companion '26), June 22--25, 2026, Banff, AB, Canada}
\acmDOI{10.1145/3765611.3815064}
\acmISBN{979-8-4007-2199-1/2026/06}

\begin{document}
\maketitle

\section{Introduction}
\label{sec:intro}

Generative-AI (GenAI) services are still in an early but fast-growing stage.
Recent indicators point to rapid expansion. The 2025 Stanford AI Index reports
that global private investment in generative AI reached \$33.9 billion in 2024
and that organizational use of generative AI in at least one business function
rose from 33\% to 71\% year over year~\citep{stanfordAiIndex2025}. Deloitte's
latest enterprise survey likewise finds AI deployment moving from pilot
programs toward broader organizational rollout~\citep{deloitteStateAi2026}.
Service prices remain high for routine use, while providers compete primarily on
model capability and quality of service. Additionally, infrastructure
deployment is driven more by frontier performance than by transparent cost
recovery. While that posture is natural in an immature sector, it is unlikely
to be the end state. When GenAI becomes a routine input to daily work and
consumption, the market scale will grow substantially and the economics of
operating the underlying infrastructure will become as important as the models
themselves.

In that regime, the relevant physical system is not a single cluster but a
geographically distributed collection of data centers, backbone links, and
service endpoints. To avoid the ambiguity of ``AI network,'' we refer to this
coupled system as \emph{distributed AI service infrastructure}. A user request
may be served by one of many candidate sites, and the marginal cost of serving
that request depends jointly on local compute availability, electricity price,
network transfer cost, and latency. This makes the allocation of AI workloads a
network-constrained resource-allocation problem rather than a purely local
server-scheduling problem.

Heterogeneity is central rather than incidental. Workload classes differ in
latency tolerance, data volume per million tokens, and energy intensity, while compute assets differ
in accelerator type, vintage, efficiency, and cost. As a result, hardware that
is no longer cost-effective for frontier interactive services can still remain
economically valuable for lower-priority, batch, or less latency-sensitive
workloads. A credible market design therefore should not assume a homogeneous
fleet. It should price a diversified asset base and reveal when older or
cheaper resources remain the efficient marginal supplier.

This observation matters for both operations and market design. Electricity is
already a major component of data-center operating expenditure, often on the
order of 20--40\% of total cost~\citep{iaei2025datacenterElectricity,networkinstallers2026costs},
while public initiatives on compute interconnection increasingly treat
distributed compute as an infrastructure resource that should be shared and
scheduled across regions~\citep{digitalchina2024supercomputingInternet,miit2025computingInterconnection,miit2026nodeNotice}.
Yet most current API pricing remains largely location-agnostic, even though the
underlying physical system is strongly location dependent. The analogy to
electricity markets should not be overstated, but the historical lesson is
relevant: once a networked service becomes critical infrastructure, efficient
dispatch and transparent locational pricing become increasingly important. We
do not claim that today's GenAI sector already operates as such a market.
Rather, we study the pricing objects and clearing rules that become relevant as
the sector matures toward one.

To this end, this paper proposes a market-clearing abstraction for that setting. We model
GenAI service provision on distributed AI service infrastructure as a
\emph{token-flow network} in which user demand is routed across communication
links and processed at compute nodes. The resulting optimization resembles
minimum-cost flow, but the economic interpretation is closer to locational
marginal pricing in electricity markets, where dual variables on the nodal balance constraints become location- and workload-specific marginal service prices.

\subsection{Background and relation to prior work}
\label{sec:related}

The paper connects three mature modeling templates. Electricity markets
show how a centrally cleared, network-constrained dispatch problem can produce
locational prices. Transportation and logistics show how commodities can be
routed through capacitated networks. Communication-network pricing shows how
link shadow prices can be interpreted as congestion signals. Our contribution is
to adapt these ideas to GenAI service tokens, where compute, transfer, and
latency are all workload dependent.

In power systems, optimal power flow and locational marginal pricing
combine nodal balance, generator limits, and line constraints to obtain prices
that decompose into energy and congestion components~\citep{frank2016opf,baldick2018lmp,caiso2010lmpappendix}.
We inherit the use of balance-constraint dual variables as marginal prices, but
our nodes are compute locations rather than buses, and the scarce local resource
is class-specific AI serving capacity rather than generation output.

Minimum-cost and multicommodity-flow models provide the mathematical
backbone for routing multiple commodities over shared capacities~\citep{bradley1977networkmodels,ahuja2001minimumcostflow,barnhart2008multicommodity,sheffi1985urban}.
The token-flow program is deliberately close to this family. The difference is not
the algorithmic structure, but the commodity definition: useful GenAI tokens
carry class-specific energy intensity, data volume per million tokens,
latency eligibility, and settlement interpretation.

Communication-network congestion control and network utility
maximization provide a complementary view in which link prices decentralize
rate-allocation decisions~\citep{kelly1998ratecontrol,low1999optimization,palomar2006decomposition}.
Those models usually price communication capacity itself. Here communication is
only one layer of the service: routing a request also consumes scarce compute at
the destination, so the economically relevant price must include both transfer
scarcity and compute scarcity.

Finally, the cloud and AI-infrastructure context motivates why these
pricing objects matter now. Data-center operating cost, public compute
interconnection initiatives, and AI-service metering all point toward a more
infrastructure-like service layer~\citep{networkinstallers2026costs,digitalchina2024supercomputingInternet,miit2025computingInterconnection}.
The novelty claimed here is therefore the
identification of the economically relevant commodity, constraints, and dual
prices for geographically distributed GenAI service infrastructure than a new flow algorithm.

\paragraph{Contributions.}
Our contributions are threefold.
\begin{itemize}
\item We formulate a linear market clearing model for GenAI services
that jointly route and process requests under compute-capacity and
bandwidth constraints on a geographically distributed and heterogeneous
infrastructure.
\item We derive the associated locational marginal service prices and show how
they decompose into energy cost, compute scarcity rent, routing cost, and link
congestion rent. We also introduce a transfer-aware extension that prices
network usage in physical data units (GB/s).
\item On 5-node and 20-node U.S. data-center networks, the case studies show
merit-order dispatch, scarcity transitions in image generation, transfer
bottlenecks hidden by token-equivalent bandwidth models, and strong regional
price separation under tight latency constraints.
\end{itemize}

The rest of the paper is organized as follows. Section~\ref{sec:model} presents
the token-flow market model and its price interpretation. Section~\ref{sec:setup}
describes the data sources, scenario construction, and calibration choices.
Section~\ref{sec:results} reports the numerical results, and
Sections~\ref{sec:discussion} and~\ref{sec:conclusion} discuss limitations and
summarize the paper.

\section{Token-Flow Market Model}
\label{sec:model}

\subsection{System abstraction}

Let $\mathcal{N}$ be the set of compute nodes and
$\mathcal{E} \subseteq \mathcal{N}\times\mathcal{N}$ the directed
communication network. Workloads are partitioned into job classes
$k \in \mathcal{K}$, such as interactive chat, image generation, code review,
and batch training. For each class $k$, demand $d_{j,k}$ denotes the useful
token rate requested at node $j$.
Operationally, $d_{j,k}$ is an exogenous request-arrival rate at origin
region $j$. The market clearing problem decides whether that demand is processed
locally or routed to another compute node subject to constraints. The injection pattern is therefore an input to the market, and changing the demand vector can change both dispatch and prices.

We use two decision variables for each class $k$:
\begin{itemize}
\item $f_{ij,k} \ge 0$: token flow of class $k$ routed on arc $(i,j)$;
\item $x_{j,k} \ge 0$: token rate of class $k$ processed at node $j$.
\end{itemize}
Each node $j$ has class-specific processing capacity $C_{j,k}$, local
electricity price $p_j^{\mathrm{elec}}$ (\$/kWh), and class-specific energy
requirement $e_{j,k}$ (kWh/M tokens). The resulting marginal processing cost is
\[
g_{j,k} := p_j^{\mathrm{elec}} e_{j,k} \qquad (\$/\text{M tokens}).
\]
Each arc $(i,j)$ has token-equivalent capacity $B_{ij}$ and routing cost
$c_{ij,k}$, which captures latency penalties, network charges, or other
per-token delivery costs in the baseline model.
For latency-sensitive workloads we restrict routing to a feasible arc set
$\mathcal{E}_k \subseteq \mathcal{E}$ defined by a class-specific latency rule.
In the numerical study, this rule is implemented by removing arcs whose latency
exceeds the class-specific threshold.

\subsection{Baseline market-clearing program}

Let $\bb{A} \in \{-1,0,1\}^{|\mathcal{N}|\times|\mathcal{E}|}$ be the
node--arc incidence matrix, with $+1$ at the sending node and $-1$ at the
receiving node for each arc, so that $\bb{A}\bb{f}_k$ is net outbound flow. For each
class $k$, let
$\bb{f}_k \in \mathbb{R}_+^{|\mathcal{E}|}$ collect arc flows,
$\bb{x}_k,\bb{d}_k,\bb{g}_k,\bb{C}_k \in \mathbb{R}_+^{|\mathcal{N}|}$ collect
node variables and parameters, and let $\bb{B}$ denote the vector of
token-equivalent arc capacities. The baseline clearing problem is formulated as the following linear program:
\begin{subequations}
\label{eq:baseline_lp}
\begin{align}
\min_{\{\bb{f}_k,\bb{x}_k\}_{k\in\mathcal{K}}}\quad &
\sum_{k\in\mathcal{K}}
\left(\bb{c}_k^\top \bb{f}_k + \bb{g}_k^\top \bb{x}_k\right)
\label{eq:baseline_lp_obj}\\
\text{s.t.}\quad &
\bb{A}\bb{f}_k + \bb{x}_k = \bb{d}_k,
\quad \forall k \in \mathcal{K},
\label{eq:baseline_lp_balance}\\
&
\bb{0} \le \bb{x}_k \le \bb{C}_k,
\quad \forall k \in \mathcal{K},
\label{eq:baseline_lp_compute}\\
&
\sum_{k\in\mathcal{K}} \bb{f}_k \le \bb{B},
\label{eq:baseline_lp_bandwidth}\\
&
\bb{f}_k \ge \bb{0},
\quad \forall k \in \mathcal{K}.
\label{eq:baseline_lp_nonneg}
\end{align}
\end{subequations}
where~\eqref{eq:baseline_lp_balance} enforces flow conservation where all demand
must either be processed locally or routed elsewhere for service.
Constraint~\eqref{eq:baseline_lp_compute} imposes node-level compute limits, and
\eqref{eq:baseline_lp_bandwidth} captures shared communication capacity. The
objective minimizes operating cost while serving all demand.

In scalar form, the balance equation is
$\sum_h f_{jh,k}+x_{j,k}=d_{j,k}+\sum_i f_{ij,k}$: origin demand and inbound
traffic must be either processed at node $j$ or sent onward. The routing
coefficient can be written as $c_{ij,k}=c^{\mathrm{route}}_{ij,k}+w_{ij}a_k$,
where $w_{ij}$ is a transfer tariff in \$/GB and $a_k$ is the payload size of a
useful token. The baseline token-equivalent bandwidth approximation sets
$B_{ij}=W_{ij}/\underline a$, with
$\underline a=\min_{k\in\mathcal K}a_k$; this keeps the LP compact but is
optimistic for data-heavy classes. Latency requirements are enforced by setting
$f_{ij,k}=0$ when $\tau_{ij}>L_k^{\max}$. Taken together, the equations define a
linear clearing problem in which electricity cost sets the merit order, node
capacity creates compute scarcity, and link limits create network scarcity.

The equality in~\eqref{eq:baseline_lp_balance} is a must-serve
abstraction. This is appropriate for a first market-clearing model under a
service-obligation interpretation, but production systems can queue, degrade, or
drop requests. A partial-service extension would introduce unmet demand
$y_{j,k}\ge 0$ and penalty $\rho_{j,k}$, replace the balance with
$\bb{A}\bb{f}_k+\bb{x}_k+\bb{y}_k=\bb{d}_k$, and add
$\sum_{j,k}\rho_{j,k}y_{j,k}$ to the objective. That extension converts
infeasibility into deferred or shed service and caps scarcity prices at the
value of lost service.

\begin{remark}[Economic-Dispatch Benchmark]
A simpler benchmark would clear demand against node-level marginal
processing costs and capacities without explicit network constraints, in the
spirit of economic dispatch in power systems~\citep{schweppe1988spot,stoft2002power}.
We use the network-constrained model as the main object because transfer and
latency are not secondary refinements for GenAI services, as they materially affect
both dispatch and prices. Section~\ref{sec:results} therefore compares nested
formulations of the same clearing abstraction rather than unrelated cloud
schedulers.
\end{remark}

\subsection{Dual prices and economic interpretation}
\label{sec:prices}

Let $\bb{\pi}_k$ be the dual multipliers on the nodal balance constraints
\eqref{eq:baseline_lp_balance}, $\bb{\mu}_k \ge \bb{0}$ the multipliers on the
compute-capacity upper bounds \eqref{eq:baseline_lp_compute}, and
$\bb{\eta} \ge \bb{0}$ the multipliers on the link-capacity constraints
\eqref{eq:baseline_lp_bandwidth}. Using the sign convention that $\bb{\pi}_k$
measures the marginal cost of one additional unit of demand, the Lagrangian of
\eqref{eq:baseline_lp} is
\begin{align}
\mathcal{L}
={}&
\sum_{k\in\mathcal{K}}
\Big[
\bb{c}_k^\top \bb{f}_k
+ \bb{g}_k^\top \bb{x}_k
- \bb{\pi}_k^\top(\bb{A}\bb{f}_k + \bb{x}_k - \bb{d}_k)
\notag\\
&\qquad\qquad
+ \bb{\mu}_k^\top(\bb{x}_k - \bb{C}_k)
\Big]
+ \bb{\eta}^\top
\left(\sum_{k\in\mathcal{K}}\bb{f}_k - \bb{B}\right).
\label{eq:baseline_lagrangian}
\end{align}
Minimizing \eqref{eq:baseline_lagrangian} over the nonnegative primal variables
produces the dual feasibility conditions
\begin{align}
\bb{A}^\top \bb{\pi}_k &\le \bb{c}_k + \bb{\eta},
\quad \forall k \in \mathcal{K},
\label{eq:dual_flow}\\
\bb{\pi}_k &\le \bb{g}_k + \bb{\mu}_k,
\quad \forall k \in \mathcal{K}.
\label{eq:dual_compute}
\end{align}
and the KKT relations
\begin{align}
0 \le \bb{c}_k + \bb{\eta} - \bb{A}^\top \bb{\pi}_k &\perp \bb{f}_k \ge 0,
\label{eq:kkt_flow}\\
0 \le \bb{g}_k + \bb{\mu}_k - \bb{\pi}_k &\perp \bb{x}_k \ge 0.
\label{eq:kkt_compute}
\end{align}

The multiplier $\pi_{j,k}$ is the marginal system cost of serving one
additional token of class $k$ at node $j$, i.e., the \emph{locational marginal
service price}. Complementary slackness gives two key identities.

For any active arc $(i,j)$ used by class $k$,
\begin{align}
\pi_{i,k}^* - \pi_{j,k}^*
= c_{ij,k} + \eta_{ij}^*.
\label{eq:arc_price}
\end{align}
Hence the price difference across a used link equals private routing cost plus a
congestion rent if the link is scarce.

For any active processing decision at node $j$,
\begin{align}
\pi_{j,k}^* = g_{j,k} + \mu_{j,k}^*
= p_j^{\mathrm{elec}} e_{j,k} + \mu_{j,k}^*.
\label{eq:node_price}
\end{align}
Thus the local service price decomposes into marginal energy cost plus a
compute-scarcity rent. If node $j$ has spare class-$k$ capacity, then
$\mu_{j,k}^*=0$.

Combining~\eqref{eq:arc_price} and~\eqref{eq:node_price} along an active path
$P(o,s)$ from origin $o$ to serving node $s$ yields
\begin{align}
\pi_{o,k}^*
= g_{s,k} + \mu_{s,k}^*
+ \sum_{(i,j)\in P(o,s)} \left(c_{ij,k} + \eta_{ij}^*\right).
\label{eq:lmp_decomposition}
\end{align}
Equation~\eqref{eq:lmp_decomposition} is the central pricing result of the
paper: token prices inherit the same energy-plus-congestion logic that underlies
electricity LMPs, but with an additional compute-scarcity term that is specific
to AI infrastructure.

\subsection{Transfer-aware extension and settlement}

The baseline model measures bandwidth in token-equivalent units. That is
insufficient when workload classes induce very different data volumes per useful
token. Let $\gamma_{ij,k}$ denote the effective data requirement (GB/token) of
serving one useful token of class $k$ on arc $(i,j)$ after protocol overhead,
let $w_{ij}$ be the exogenous transfer tariff (\$/GB), and let $W_{ij}$ be the
physical link capacity (GB/s). The transfer-aware refinement becomes
\begin{subequations}
\label{eq:transfer_lp}
\begin{align}
\min_{\{\bb{f}_k,\bb{x}_k\}_{k\in\mathcal{K}}}\quad &
\sum_{k\in\mathcal{K}}
\left((\bb{c}^{\mathrm{route}}_k)^\top \bb{f}_k
+ \bb{g}_k^\top \bb{x}_k
+ \bb{w}^\top \bb{\Gamma}_k \bb{f}_k\right)
\label{eq:transfer_lp_obj}\\
\text{s.t.}\quad &
\bb{A}\bb{f}_k + \bb{x}_k = \bb{d}_k,
\quad \forall k \in \mathcal{K},
\label{eq:transfer_lp_balance}\\
&
\bb{0} \le \bb{x}_k \le \bb{C}_k,
\quad \forall k \in \mathcal{K},
\label{eq:transfer_lp_compute}\\
&
\sum_{k\in\mathcal{K}} \bb{\Gamma}_k \bb{f}_k \le \bb{W},
\label{eq:transfer_lp_bandwidth}\\
&
\bb{f}_k \ge \bb{0},
\quad \forall k \in \mathcal{K},
\label{eq:transfer_lp_nonneg}
\end{align}
\end{subequations}
where $\bb{\Gamma}_k := \operatorname{diag}(\gamma_{ij,k})$ over arcs
$(i,j)\in\mathcal E$.

For any active transfer arc, the dual relation becomes
\begin{align}
\pi_{i,k}^* - \pi_{j,k}^*
= c^{\mathrm{route}}_{ij,k} + \gamma_{ij,k}\left(w_{ij} + \eta_{ij}^*\right).
\label{eq:transfer_price}
\end{align}
This version isolates physical-network payments and gives the congestion rent
units of \$/GB instead of \$/token.

The same dual variables define an economically consistent settlement rule.
Demand at node $j$ pays $\pi_{j,k}^* d_{j,k}$, compute providers receive
$(g_{j,k}+\mu_{j,k}^*)x_{j,k}^*$, and network providers receive
$(w_{ij}+\eta_{ij}^*) q_{ij,k}$ on carried traffic
$q_{ij,k} = \gamma_{ij,k} f_{ij,k}$. Any residual is a merchandising surplus
analogous to congestion revenue in electricity markets.

\section{Experimental Setup}
\label{sec:setup}

For case studies we aim to evaluate three questions: (i) what dispatch and price patterns emerge in the
baseline token-flow market, (ii) how scarcity and congestion evolve as demand
grows, and (iii) how transfer-aware and latency-aware variants change the market
outcome.

All experiments are generated with the \texttt{TokenFlowMarket.jl} artifact\footnote{Code and examples available at \url{https://github.com/ShaohuiLiu/TokenFlowMarket.git}.},
implemented in JuMP with Gurobi as the default LP/QP solver and with explicit
dual extraction for nodal prices, link congestion rents, and compute scarcity
rents. The primary dataset is a 20-hub U.S. registry compiled from Data Center
Map and manually curated into a table containing hub identifier, metro name,
state, latitude, longitude, facility count, and estimated site power
(MW)~\citep{datacentermap2026usa}. State-level electricity prices are taken
from 2025 EIA industrial retail rates and joined by
state~\citep{eia2026retailsales}. The 5-node testbed is the subset
Ashburn/NoVA, Dallas/Fort Worth, Silicon Valley, Chicago, and
Seattle/Quincy; the 20-node experiment uses the full registry.

Scenario construction follows the package's public data pipeline. First,
estimated site power is converted into synthetic class-specific processing capacities using
fixed throughput-per-MW coefficients: 5000 chatbot tokens/s/MW, 200 image
tokens/s/MW, 4000 code-review tokens/s/MW, and 1000 batch-training
tokens/s/MW. Second, bidirectional arcs are created between hubs within
2500~km, while a synthetic tier-1 backbone connects the major metros
Ashburn, Dallas, Silicon Valley, Chicago, Atlanta, and New York at
100~Gbps; all other links use 10~Gbps. Link latency is approximated as fiber
propagation plus a 5~ms per-hop overhead,
$\tau_{ij} = \mathrm{dist}_{ij}/200 + 5$, and the transfer tariff is fixed at
\$0.01/GB. Third, demand is generated from state-level population weights and
base per-class rates of 50{,}000 chatbot, 5{,}000 image-generation, 30{,}000
code-review, and 10{,}000 batch-training tokens/s at weight~1.0. The primary
operating point uses demand scale $35\times$, which yields 16.29 million
tokens/s of aggregate demand and places the system in an economically
interesting interior regime with both compute scarcity and network congestion.
We also report a 20-node U.S. scale-up to test whether the same qualitative
patterns persist in a larger network.

Note that these throughput, energy, and transfer-volume coefficients are stylized
scenario parameters rather than measurements from a particular provider. They
are chosen to separate workload classes by plausible orders of magnitude, where image
generation is compute- and transfer-heavy, chat is latency-sensitive and
data-light, and batch work is more delay tolerant. The precise scarcity
thresholds should therefore be read as scenario outcomes, while the qualitative
mechanisms rely on relative heterogeneity. While we hold explicit
throughput and transfer-volume sensitivity for future work, the present paper reports demand, latency, and electricity-price sweeps.

\begin{table}[t]
\centering
\caption{Five-node testbed used for the case study.}
\label{tab:testbed_sites}
\footnotesize
\begin{tabular}{@{}lrr@{}}
\toprule
Site & Power (MW) & Elec. price (\$/kWh) \\
\midrule
Ashburn/NoVA & 2000 & 0.078 \\
Dallas/Fort Worth & 1200 & 0.072 \\
Silicon Valley & 800 & 0.180 \\
Chicago & 1000 & 0.085 \\
Seattle/Quincy & 600 & 0.052 \\
\bottomrule
\end{tabular}
\end{table}

Table~\ref{tab:workloads} summarizes the four typical workload classes using
per-million-token units for readability. They differ by more than two orders of
magnitude in energy intensity and by roughly five orders of magnitude in data
volume, which is precisely why pricing compute and transfer jointly is
important.

\begin{table}[t]
\centering
\caption{Workload classes used in the case study in per million useful tokens.}
\label{tab:workloads}
\footnotesize
\begin{tabular}{@{}lccc@{}}
\toprule
Class & GB/M tokens & kWh/M tokens & Latency bound \\
\midrule
Interactive chat & 0.5 & 1 & 100 ms \\
Image generation & 100 & 500 & unconstrained \\
Code review & 1 & 2 & 200 ms \\
Batch training & 10 & 10 & unconstrained \\
\bottomrule
\end{tabular}
\end{table}

\section{Numerical Results}
\label{sec:results}

\begin{figure*}[t]
\centering
\includegraphics[width=0.88\textwidth]{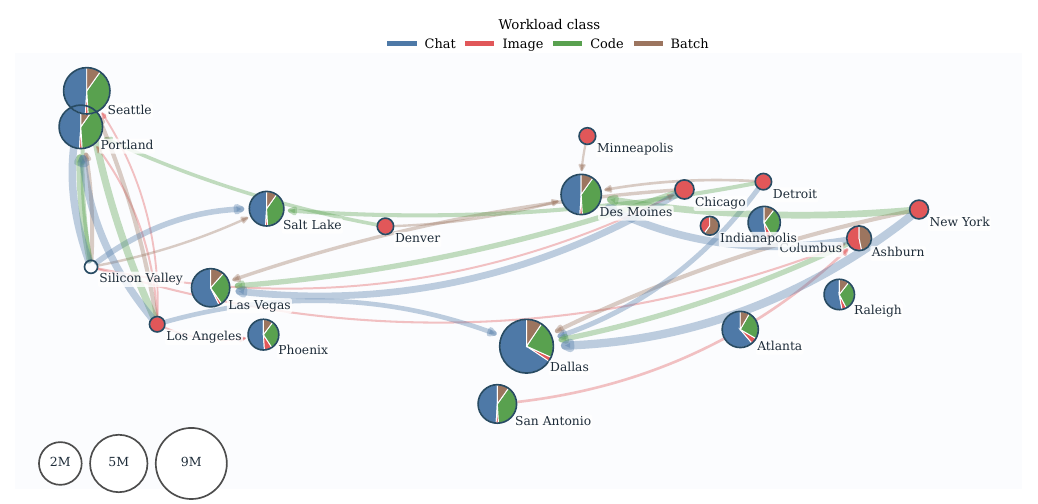}
\caption{20-node U.S. scale-up at demand scale $35\times$. Node pies show the
processing mix across chat, image, code-review, and batch-training workloads,
with pie size proportional to total processing. Colored curves show the largest
active inter-node transfers for each workload class, with thicker curves
indicating higher transferred token volume. The figure makes the merit-order
geography visible: low-cost hubs in the Northwest, Midwest, and Texas absorb
most chatbot traffic, while multiple workload classes share the same backbone
routes.}
\Description{A full-width geographic map of twenty U.S. data-center hubs. Each
node is drawn as a pie chart whose slices represent chat, image generation,
code review, and batch training processing, and larger pies indicate more total
processing. Colored curved arrows show the largest active inter-node transfers
for each workload class, with thicker arrows indicating larger transferred
volume. Large nodes appear in the Pacific Northwest, Midwest, and Texas, and
multiple classes share the main east-west transfer corridors.}
\label{fig:fullscale}
\end{figure*}

\subsection{Baseline dispatch follows locational cost and capacity scarcity}

\begin{figure*}[t]
\centering
\includegraphics[width=0.8\textwidth]{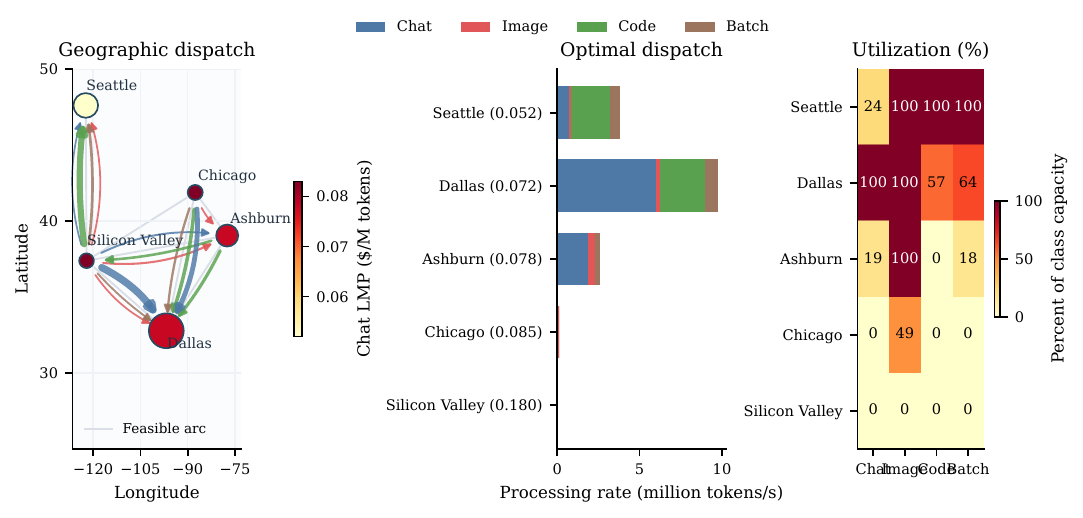}
\caption{5-node market outcome at demand scale $35\times$. Left:
geographic dispatch map with node size proportional to total processing and
active flows drawn between metros and colored by workload class. Center: optimal processing by node and
workload class; parenthesized node labels report electricity prices in \$/kWh. Right: class-specific utilization rates. Cheap nodes carry most
of the system load, while image generation and code-review capacity become
locally scarce.}
\Description{A three-panel figure for the 5-node baseline case. The left panel
is a geographic network map of the continental United States with five hubs;
node size indicates total processing, node color indicates chatbot price, and
active flows are drawn between metros. The center panel is a stacked horizontal
bar chart of processing volume by node and workload class; Dallas and Seattle
carry most of the load, while Silicon Valley is near zero. The right panel is a
heatmap of utilization percentages by node and workload class, with image
generation and code review binding at several low-cost nodes.}
\label{fig:baseline}
\end{figure*}

Figure~\ref{fig:baseline} combines a geographic and algebraic view of the
5-node baseline clearing outcome at the primary operating point. The key dispatch pattern
is in merit-order that lower-cost nodes absorb most of the workload, while
the most expensive node is displaced. Dallas and Seattle serve as the primary
workhorses, whereas Silicon Valley is completely priced out despite nontrivial installed capacity. The result shows direct implication of nodal prices being anchored to the cheapest feasible marginal supplier.

The pricing decomposition confirms that scarcity rents are concentrated in the classes and locations where capacity binds. Table~\ref{tab:price_decomp} reports selected node--class pairs, not the full site list. Chicago carries image-generation volume in Figure~\ref{fig:baseline}, and Silicon Valley is part of the network but is displaced at this operating point. Seattle's image-generation price is especially revealing: only \$26.00/M tokens comes from energy, while \$15.47/M tokens, or 37.3\% of the total price, comes from compute scarcity. By contrast, Ashburn's chatbot price is entirely explained by energy cost because the corresponding capacity is slack.

\begin{table}[t]
\centering
\caption{Selected locational price decompositions at the 5-node baseline
operating point reported in \$/M tokens where $\text{LMP} = \text{Energy cost} + \text{Scarcity charge}$.  }
\label{tab:price_decomp}
\footnotesize
\begin{tabular}{@{}llrrr@{}}
\toprule
Node & Class & LMP & Energy & Scarcity \\
\midrule
Ashburn & Chat & 0.078 & 0.078 & 0 \\
Dallas & Chat & 0.078 & 0.072 & 0.006 \\
Dallas & Image & 41.50 & 36.00 & 5.50 \\
Seattle & Image & 41.47 & 26.00 & 15.47 \\
Seattle & Batch & 0.694 & 0.520 & 0.174 \\
\bottomrule
\end{tabular}
\end{table}

Note that one subtle but important implication is that a node's local energy cost does not
by itself determine its price. Silicon Valley's chatbot energy cost is
\$0.18/M tokens, but because no chatbot demand is served there in the baseline
optimum, the local nodal price is instead pinned down by the cheapest feasible
external supply chain. In other words, locational marginal prices are
system-level marginal values, not local average costs.

\subsection{Sharp scarcity transition from demand growth}

\begin{figure*}[t]
\centering
\includegraphics[width=0.9\textwidth]{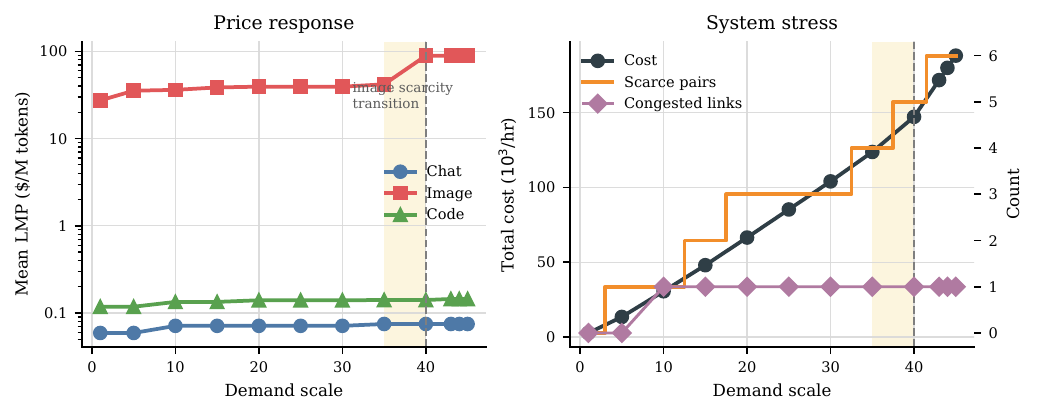}
\caption{Market dynamics under demand growth in the 5-node network. Left:
mean locational prices by workload class in \$/M tokens. Right: total operating cost, scarce
node--class pairs, and congested links. The most pronounced transition occurs in
image generation between scales $35\times$ and $40\times$; the log price axis
compresses the visual jump.}
\Description{A two-panel line chart over demand scale. The left panel shows mean
locational prices for chat, image generation, and code review; the image line
jumps sharply near scale forty. The right panel shows total system cost, the
number of scarce node-class pairs, and the number of congested links; both
system stress measures rise with demand.}
\label{fig:scaling}
\end{figure*}

Next we sweep the demand scale from $1\times$ to $45\times$, as in Figure~\ref{fig:scaling}, and two features stand out. First, mean chatbot prices are relatively stable because chatbot capacity remains abundant over most of the range. Second, image
generation exhibits a pronounced scarcity cliff. Between scale $35\times$ and
$40\times$, the mean image-generation price rises from \$41.9/M tokens to
\$89.2/M tokens, while the number of scarce node--class pairs grows
from four to five. This is the same qualitative behavior observed in many
electricity markets near peak load, where prices remain moderate until a binding capacity margin disappears, and then the marginal price rises sharply.

The congestion pattern emerges earlier than the price cliff. The first congested
link appears by scale $10\times$, but system cost and image prices remain fairly
smooth until image-generation capacity becomes tight across most of the network.
This distinction is operationally important, as a planner can observe congestion well before the market reaches the more severe regime in which small demand increases generate large price changes.

\subsection{Transfer-aware and latency-aware models reveal hidden bottlenecks}

Table~\ref{tab:formulation_compare} treats the main variants as nested
abstractions rather than unrelated algorithms. The transfer-aware model in
\eqref{eq:transfer_lp} changes the economic interpretation of the network
constraints in a substantive way. Using the baseline LP formulation, the 5-node system
reports one congested token-equivalent link and a cost of \$123{,}635/hr. Under
the transfer-aware LP, cost rises to \$127{,}007/hr with four links saturate
physically. The difference is driven almost entirely by image generation, whose
bytes-per-token ratio is large enough to consume entire backbone links even when
the useful token flow looks modest.

\begin{table}[t]
\centering
\caption{Nested formulation comparison at the 5-node, $35\times$ operating point. Mean prices are reported in \$/M tokens.}
\label{tab:formulation_compare}
\footnotesize
\begin{tabular}{@{}lrrrr@{}}
\toprule
Model & Cost/hr & Chat & Image & Congestion \\
\midrule
Baseline LP & \$123{,}635 & 0.0748 & 41.90 & 1 \\
Transfer-aware LP & \$127{,}007 & 0.0770 & 38.90 & 4 \\
Congestion-QP variant & \$123{,}635 & 0.0748 & 41.90 & 1 \\
Uniform Opex adder & \$163{,}140 & 0.3186 & 47.99 & 1 \\
\bottomrule
\end{tabular}
\end{table}

Latency constraints change prices even more sharply. Tightening the chatbot latency requirement from 100~ms to 15~ms and the code-review limit from 200~ms to 20~ms partitions the 5-node network into an eastern cluster (Ashburn, Dallas, Chicago) and a western cluster (Silicon Valley, Seattle). While overall cost increase is modest at only 0.29\%,  the local price impact shows dramatic changes: Silicon Valley's chatbot price rises from \$0.083/M tokens to \$0.180/M tokens with a 117\% increase, due to a fact that it can no longer export latency-sensitive demand to cheaper eastern nodes.

\begin{figure*}[t]
\centering
\includegraphics[width=0.9\textwidth]{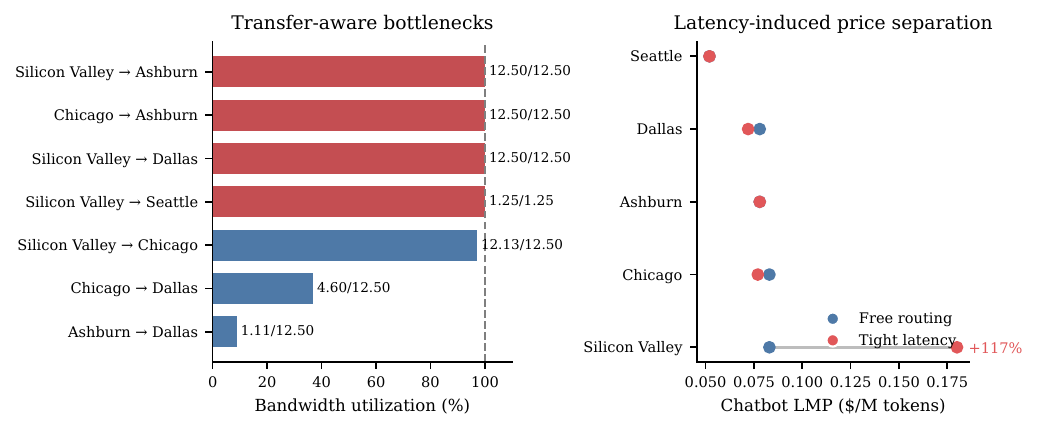}
\caption{Left: physical bandwidth utilization of
active links under the transfer-aware LP with four links fully saturated. Right:
chatbot locational prices in \$/M tokens before and after imposing tight latency limits. The
largest increase occurs at Silicon Valley, which becomes partially isolated from
cheaper eastern supply.}
\Description{A two-panel figure showing the effect of model extensions. The left
panel is a horizontal bar chart of bandwidth utilization by active link in the
transfer-aware model, with four links at full capacity. The right panel is a
dumbbell chart comparing chatbot prices before and after tight latency limits;
Silicon Valley shows the largest increase.}
\label{fig:extensions}
\end{figure*}

Note that adding the uniform non-energy
Opex adder used in the extended-cost experiment increases total cost from
\$123{,}635/hr to \$163{,}140/hr in a 31.95\% increase, but leaves the optimal
dispatch unchanged because the adder shifts all nodes in parallel. This contrast
reveals a useful insight that heterogeneous transfer and latency effects change allocations, whereas
uniform accounting adders mostly change the price level.

\subsection{Scale-up and settlement}

The same qualitative patterns persist in the 20-node network. At demand scale
$35\times$, the baseline model clears 43.89 million tokens/s at
\$353{,}702/hr with 28 scarce node--class pairs. Dispatch again follows
electricity merit order that cheap hubs such as Seattle, Portland, Des Moines, and
Dallas are heavily utilized, while expensive California nodes are displaced for
chatbot traffic and only re-enter when image-generation capacity becomes scarce.
Figure~\ref{fig:fullscale} makes the spatial pattern visible. Chatbot service
concentrates in the Pacific Northwest, Mountain West, Midwest, and Texas, while
image-generation utilization is close to 100\% at most sites. The 20-node
system becomes infeasible at scale $40\times$, indicating that aggregate
image-generation capacity, not just local congestion, becomes the dominant
system bottleneck.

Notet that the settlement implied by the dual prices is revenue adequate in the
5-node baseline case. Aggregate user payments are \$139{,}446/hr, compute
providers receive \$137{,}613/hr, network providers receive \$1{,}784/hr, and
the residual merchandising surplus is \$49.71/hr. This calculation should be
read as an accounting interpretation of the dual variables, not as an
independent empirical validation. Its value is to show how payments split across
compute, transfer, and residual congestion revenue under a nodal service-price
settlement. Tracking how the surplus changes with demand and congestion is a
natural extension for realistic market considerations.

\section{Discussion}
\label{sec:discussion}

The authors would like to emphasize that the proposed framework should be read as a model of an emerging infrastructure market rather than as a precise description of today's API business. Current providers still compete primarily on capability, latency, and ecosystem, and many pricing decisions are strategic rather than marginal-cost based in this emerging sector with (nearly) sustaining capital inflows. That is typical of an early-growth sector. The value of the present formulation is to identify the pricing objects that become important once adoption broadens, infrastructure utilization tightens, and return on geographically distributed assets becomes an operational concern.

The model is especially useful in a heterogeneous fleet. Different accelerators, hardware vintages, and service classes imply different throughput, energy intensity, transfer demand, and latency feasibility. That is exactly why older hardware (such as Nvidia A100, H100) can remain valuable for lower-priority or batch services even after it ceases to be competitive for frontier interactive workloads. A market that exposes nodal compute scarcity and transfer scarcity can therefore improve short-run dispatch efficiency and, eventually, provide clearer signals for specialization, interconnection, and asset-mix decisions.

The results should therefore not be reduced to the simple rule ``put data
centers where electricity is cheap.'' Low electricity price is the first-order
merit-order driver, but it is not sufficient for the fact that transfer-heavy workloads can
saturate links, and latency-sensitive workloads may require nearby capacity even
when local electricity is expensive. Remote low-cost regions are naturally
valuable for delay-tolerant or data-light workloads; expensive urban or coastal
capacity can still have option value for interactive services; and
interconnection investment has a measurable shadow value through link congestion
rents. Consumer-facing prices also need not expose raw locational volatility like wholesale electricity prices. Longer-term
contracts, subscriptions, or two-part tariffs could smooth marginal service
costs in the same spirit as retail electricity contracts. The point is that the underlying marginal prices are still important for operational control and investment decisions, even if they are not directly visible to end users.

Lastly, the formulation is intentionally stylized. It is static, deterministic, and continuous. And it does not represent queueing dynamics, discrete accelerator allocation, reliability constraints, carbon accounting, or privacy restrictions on data movement. It also treats the power system as exogenous, whereas in practice the AI service infrastructure and electricity system are coupled through data-center load. These simplifications matter if the goal is operational deployment. They are less problematic if the goal is to identify the right marginal pricing objects. For that purpose, the model isolates the essential economics cleanly, where energy cost determines the baseline merit order, compute scarcity adds nodal rents, and data transport adds location-specific congestion rents that depend on workload characteristics. Our ongoing work extends the framework to multi-period demand uncertainty, explicit investment decisions, hardware retirement and repurposing, and joint optimization with the underlying power grid.

\section{Conclusion}
\label{sec:conclusion}

We presented a network-constrained token-flow market for distributed GenAI services on geographically distributed AI service infrastructure. The model yields locational marginal service prices that decompose into energy, scarcity, and transfer components, while the transfer-aware extension prices network usage in physical data units.

The numerical study shows that the prices identify merit-order dispatch, expose capacity cliffs, reveal bandwidth bottlenecks, and quantify the regional market segmentation created by tight latency requirements. This suggests that locational pricing is a useful design principle for an emerging AI service infrastructure in which heterogeneous workloads and heterogeneous assets must be coordinated rather than treated as an undifferentiated cloud resource. If GenAI continues to diffuse into daily life and work, such pricing mechanisms are likely to become more important, not less, for both operational control and the design of future competitive markets.

\section*{Acknowledgement}
Shaohui Liu is partially supported by the MIT Energy Initiative Future Energy System Center. The experiments are partially supported by U.S. National Science Foundation ACCESS and NAIRR programs. The authors thank Dr. Shen Wang, Dr. Audun Botterud, and the anonymous reviewers for their constructive feedback and suggestions.
The authors acknowledge the use of ChatGPT and Claude to assist in the writing and editing of this manuscript. The authors reviewed and approved all content for accuracy and originality.

\bibliographystyle{ACM-Reference-Format}
\bibliography{token_market_refs}

\end{document}